\documentclass[usenatbib]{mn2e}
\usepackage{graphicx}

\title[On the t$_2$ - t$_3$ relation]
      {On the relation between electron temperatures in the O$^{+}$ and 
       O$^{++}$ zones in high-metallicity H\,{\sc ii} regions}
      
\author[L.S.Pilyugin, J.M.V\'{\i}lchez and T.X.Thuan]
       {Leonid S. Pilyugin$^{1}$, Jos\'{e} M. V\'{\i}lchez$^{2}$ 
       and Trinh X.Thuan$^{3}$\\
       $^{1}$Main Astronomical Observatory
             of National Academy of Sciences of Ukraine,
             27 Zabolotnogo str., 03680 Kiev, Ukraine
             (pilyugin@mao.kiev.ua)              \\
       $^{2}$Instituto de Astrof\'{\i}sica de Andaluc\'{\i}a,
             CSIC, Apdo, 3004, 18080 Granada, Spain
             (jvm@iaa.es)              \\
       $^{3}$Astronomy Department, University of Virginia, P.O.Box 3818, 
             University Station, Charlottesville, VA 22903
             (txt@virginia.edu)         \\}

\date{Accepted 2006 May 25. Received 2006 May 25; in original form 2006 April 13}

\begin{document}

\maketitle

\begin{abstract}
We suggest a new way to establish the 
relation between the electron temperature 
$t_3$ within the [O\,{\sc iii}] zone and the electron temperature $t_2$ within 
the [O\,{\sc ii}] zone in high-metallicity (12+log(O/H) $>$ 8.25) H\,{\sc ii} 
regions. The t$_2$ -- t$_3$ diagram is constructed by applying our 
method to a sample of 372 H\,{\sc ii} regions.  We find that the 
correlation between t$_2$ and t$_3$ is tight and can be approximated by a 
linear expression. The new $t_2$ -- $t_3$ relation can be used to 
determine t$_2$ and accurate abundances in high-metallicity 
H\,{\sc ii} regions with a measured t$_3$. It can also be  
used in conjunction with the ff relation for the determination of t$_3$ 
and t$_2$ and oxygen abundances in high-metallicity H\,{\sc ii} 
regions where the [OIII]$\lambda$4363 auroral line is not detected. 
The derived $t_2$ -- $t_3$ relation is independent of photoionization models 
of H\,{\sc ii} regions.
\end{abstract}


\section{Introduction}

Accurate abundances in H\,{\sc ii} regions can be derived via the classic
T$_{\rm e}$  method, T$_{\rm e}$ being the electron temperature of the  
H\,{\sc ii} region. In this method, the reliability of the equations 
for determining the O$^{++}$/H$^+$ and O$^{+}$/H$^+$ ionic oxygen abundances 
depends to a great part on the reliability of the atomic data.
\citet{pilyuginthuan05}  have compared the ionic oxygen abundances 
derived with a recent set of equations using the latest atomic data 
(\citet{izotovetal05}), with those determined from an earlier set of 
equations (\citet{pageletal92}) and have found close agreement. 
Thus, the equations for the determination of the ionic oxygen abundances in 
H\,{\sc ii} regions appear to be robust, in the sense that, for a given set of 
measured electron temperatures t$_3$ and t$_2$, relations  
determined by different authors result in 
close oxygen abundances in H\,{\sc ii} regions.

The electron temperatures $t_3$ within the [O\,{\sc iii}] zone and $t_2$ 
within the [O\,{\sc ii}] zone are determined from diagnostic line ratios when 
they are available. The intensity ratio of the 
[O\,{\sc iii}]$\lambda 4959+\lambda 5007$ nebular line to the 
[O\,{\sc iii}]$\lambda 4363$ auroral line is used to determine t$_3$. As for 
t$_2$, the intensity ratio of the [O\,{\sc ii}]$\lambda 3727$ nebular line to 
the [O\,{\sc ii}]$\lambda 7320+\lambda 7330$ auroral line is used. Based on a 
sample of H\,{\sc ii} regions where all the necessary oxygen lines are detected, 
\citet{kennicuttetal03} found that the t$_2$ temperatures show a large scatter 
and that they are nearly uncorrelated with t$_3$ temperatures. They noted that 
the measurements of the faint [O\,{\sc ii}]$\lambda 7320+\lambda 7330$ auroral 
lines may contain large random errors. Furthermore, 
\citet{rubin86,tsamisetal03} have pointed out that the  
[O\,{\sc ii}]$\lambda 3727$/[O\,{\sc ii}]$\lambda 7320+\lambda 7330$ ratio 
may be affected by recombination. \citet{izotovetal05} have derived t$_2$ and 
t$_3$ from the above ratios for a sample of H\,{\sc ii} regions. They concluded 
that the two temperatures generally follow the t$_2$ -- t$_3$ relation obtained 
from photoionization models (e.g. Stasinska 1982, 1990), but that the scatter 
of the data points is very large. The large scatter was attributed to large 
flux errors of the weak [O\,{\sc ii}]$\lambda 7320+\lambda 7330$ emission lines.

The intensity ratio of the [N\,{\sc ii}]$\lambda 6548+\lambda 6584$ nebular line 
to the [N\,{\sc ii}]$\lambda 5755$ auroral line is also used to determine 
t$_2$. It seems to give more reliable values. However, the contribution of 
recombination to the excitation of the [N\,{\sc ii}]$\lambda 5755$ line may 
affect temperatures derived from the [N\,{\sc ii}] nebular to auroral line ratios 
\citep{rubin86,tsamisetal03}. The 
([N\,{\sc ii}]$\lambda 6548+\lambda 6584$)/[N\,{\sc ii}]$\lambda 5755$ ratio 
is measured in only a few H\,{\sc ii} regions with a detected  
([O\,{\sc iii}]$\lambda 4959+\lambda 5007$)/[O\,{\sc iii}]$\lambda 4363$ line 
ratio. 

When only a single temperature measurement is available, a t$_2$ -- t$_3$ 
relation based on grids of H\,{\sc ii} region models is usually used. 
Several versions of such a $t_2$ -- $t_3$ relation have been proposed
\citep[e.g.][]{campbelletal86,pageletal92,izotovetal97,deharvengetal00,oeyshields00}, 
but the agreement between them is not very good. The available measurements 
do not provide an undisputable evidences in favour of any out of suggested 
relations, i.e. the choice of the relation is in fact arbitrary. Thus, the 
$t_2$ -- $t_3$ relation seems to be the weakest link of the classic 
T$_{\rm e}$  method. We will address this problem here.

The main goal of this paper is to derive a t$_2$ -- t$_3$ relation which is 
model-independent. This allows one to relax the arbitrarity in the choice 
of the $t_2$ -- $t_3$ relation. 
The basic idea is the following. Usually, the set of 
equations used for the determination of the ionic O$^{++}$/H$^+$ and 
O$^{+}$/H$^+$ abundances in H\,{\sc ii} regions is applied to the whole nebula. 
If we apply the equation for O$^{++}$/H$^+$ only to the O$^{++}$ zone, 
then this would yield, not the ionic O$^{++}$/H$^+$ abundance, but the total 
O/H oxygen abundance instead. There is no reason to suspect that the oxygen 
abundances in the O$^{++}$ and O$^{+}$ zones differ. We therefore require that 
the equation for O$^{++}$/H$^+$ applied to the O$^{++}$ zone and the one for 
O$^{+}$/H$^+$ applied to the O$^{+}$ zone result in exactly the same value of 
the oxygen abundance. This condition allows us to derive a relation between 
t$_2$ and t$_3$. We can apply the equation for O$^{++}$/H$^+$ to the O$^{++}$ 
zone alone, only if we know the contribution of the O$^{++}$ zone to the 
measured H$_{\beta}$ flux of the H\,{\sc ii} region. We will be using the 
ff relation \citep{ff,pilyuginetal06} for this purpose. 

The observational sample to be used is 
described in Section 2. The strategy for determining the t$_2$ -- t$_3$ relation 
is discussed in Section 3. A model-independent t$_2$ -- t$_3$ relation is 
derived in Section 4. We discuss the results in Section 5, and summarize our 
conclusions in Section 6. 

We will be using the following notations throughout the paper:
R$_2$ = I$_{{\rm [OII] \lambda 3727+ \lambda 3729}}$/I$_{{\rm H_{\beta} }}$,
R$_3$ = I$_{{\rm [OIII] \lambda 4959+ \lambda 5007}}$/I$_{{\rm H_{\beta} }}$,
R = I$_{{\rm [OIII] \lambda 4363}}$/I$_{{\rm H_{\beta} }}$, R$_{23}$ = R$_2$ + R$_3$.
With these definitions, the excitation parameter P can be 
expressed as: P = R$_3$/(R$_2$+R$_3$).

\section{Observational data}

A large sample of high-precision measurements of H\,{\sc ii} regions
 is at the base of the present investigation. We include first the 
sample of  
\citet{pilyuginthuan05} who have carried out an extensive search of the 
literature to compile a list of more than 700 individual spectra of 
H\,{\sc ii} regions in irregular and spiral galaxies, with the requirement 
that they all possess a detected [O\,{\sc iii}]~$\lambda 4363$ emission line. 
While we have tried to include as many sources as possible, we do not claim 
our search to be exhaustive. Since the majority of extragalactic 
H\,{\sc ii} regions are in the low-density regime \citep{zkh,bresolinetal05}, 
only such H\,{\sc ii} regions will be considered here. 
Of the objects with the electron density-sensitive ratio 
r$_{\rm n}$ = [S\,{\sc ii}]~$\lambda 6716$/[S\,{\sc ii}]~$\lambda 6731$ 
available, those with r$_{\rm n}$ $<$ 1.3 were excluded.

We have also included the sample of 
\citet{izotovetal04,izotovetal05} who have 
extracted from 
the Data Release 3 of the Sloan Digital Sky Survey
(SDSS) around 4500 spectra of H\,{\sc ii} 
regions with an [O\,{\sc iii}]~$\lambda 4363$ emission line detected at 
a level better than 1$\sigma$, and carefully measured the line intensities 
in each spectrum. The SDSS  
H\,{\sc ii} regions are also in a low-density regime. 
Yuri Izotov and Natalia Guseva have kindly provided 
us with the total list of their measurements, as only 
part of these have been published \citep{izotovetal04,izotovetal05}. 
In total, our sample consists of around 5200 H\,{\sc ii}
region spectra.

Following \citet{pilyuginthuan05}, we can use the the ff  relation  
\citep{pilyuginetal06} to select out the H\,{\sc ii} regions with the 
highest quality measurements. We can define a ``discrepancy index'',
equal to the  difference between the logarithm of the 
observed flux R$^{{\rm obs}}$ in the [O\,{\sc iii}]$\lambda$4363 line 
and the logarithm of the flux R$^{\rm cal}$ of that line derived from the 
strong [O\,{\sc ii}]$\lambda$3727, [O\,{\sc iii}]$\lambda$$\lambda$4959,5007 
lines using the ff relation: 
\begin{equation}
D_{{\rm ff}} =  \log R^{{\rm obs}} - \log R^{\rm cal}.
\label{equation:dff}
\end{equation}

\begin{figure}
\resizebox{1.00\hsize}{!}{\includegraphics[angle=000]{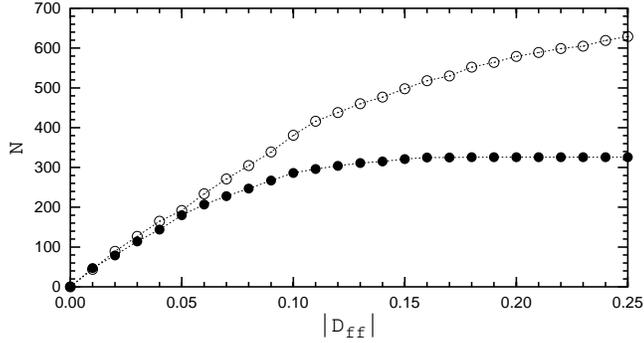}}
\caption{Cumulative number of individual measurements of H\,{\sc ii} regions 
with the absolute value of the discrepancy index D$_{\rm ff}$ less than a 
given value. The solid circles show 
data for H\,{\sc ii} regions with positive values of 
the discrepancy index, and the open circles for H\,{\sc ii} regions with 
negative values of the discrepancy index. }
\label{figure:gist}
\end{figure}

It is well known that the relation between the oxygen abundance and the 
strong oxygen line intensities is double-valued, with two distincts parts,
traditionally known as the upper high-metallicity branch 
and the lower low-metallicity branch of the R$_{23}$ -- O/H 
diagram. Following \citet{pilyuginthuan05}, we adopt the value of 12+log(O/H) 
= 8.25 as the boundary between the upper branch and the transition zone. 
The exact boundary is difficult to establish, but we have chosen that value 
because we will be using later the ff relation which is applicable 
at metallicities above 12+log(O/H) $\sim$ 8.25. We have 
extracted from our total sample a subsample of high-metallicity H\,{\sc ii} 
regions with an oxygen abundance 12+log(O/H) $>$ 8.25. The oxygen abundances 
have been calculated with the equations of \citet{izotovetal05} for the 
T$_{\rm e}$ method, and with the t$_2$ -- t$_3$ relation of \citet{campbelletal86}. 
Fig.~\ref{figure:gist} shows the cumulative number of individual H\,{\sc ii} 
region measurements with the absolute value of the discrepancy index 
D$_{\rm ff}$ less than a certain value. Solid and open circles show  
data for H\,{\sc ii} regions with respectively positive and negative values of 
the discrepancy index. It is seen that the agreement between the cumulative 
numbers of measurements with positive and negative values of the discrepancy 
index D$_{\rm ff}$ is good for values of D$_{\rm ff}$ less than 0.05, but gets 
increasingly worse at higher values. This is the result of a selection effect. 
Indeed, a large positive value of D$_{\rm ff}$ means that the observed 
R$^{{\rm obs}}$ is significantly overestimated. This leads to a substantial 
overestimate of the  electron temperature and an underestimate of the 
oxygen abundance, bringing it lower than our metallicity 
cut-off of 12+log(O/H) = 8.25. This effect thus systematically depletes
the number of high-metallicity H\,{\sc ii} regions with a large positive
D$_{\rm ff}$. To free ourselves from this systematic effect, we select only 
measurements with the absolute value of D$_{\rm ff}$ less than 0.05. This sample 
contains 372 data points and will be referred to as the standard sample. 
The H\,{\sc ii} regions in it have line intensity errors that are random.

\section{A strategy for the determination of the \lowercase{t$_2$} -- 
\lowercase{t$_3$} relation}

\subsection{The usual version of the T$_{{\rm e}}$ method}

For abundance determination in an H\,{\sc ii} 
region, a 
two-zone model for its temperature structure is usually adopted. 
\citet{izotovetal05} have recently published a set 
of equations for the determination of the oxygen abundance 
in H\,{\sc ii} regions in the context of such a two-zone model. 
According to those authors,   
the electron temperature $t_3$ within the [O\,{\sc iii}] zone, in units of 10$^4$K, 
is given by the following equation 
\begin{eqnarray}
t_3 = \frac{1.432}{\log (R_{3}/R)  - \log C_{\rm T}}    ,
\label{equation:t3}
\end{eqnarray}
The quantity  C$_{\rm T}$ is defined by: 
\begin{equation}
C_{\rm T} = (8.44 - 1.09\,t_3 + 0.5\,t_3^2 - 0.08\,t_3^3) \, v ,
\end{equation}
where
\begin{equation}
v = \frac{1 + 0.0004\,x_3}{1 + 0.044\,x_3}  ,
\end{equation}
and
\begin{equation}
x_3= 10^{-4} n_{\rm e} t_3^{-1/2}.
\end{equation}

As for the ionic oxygen abundances, they are given by the following equations: 
\begin{eqnarray}
12+ \log (O^{++}/H^+) = \log (I_{[OIII] \lambda 4959+ \lambda 5007}/I_{H_{\beta} })  +
  \nonumber  \\
6.200 + \frac{1.251}{t_3}  - 0.55 \log t_3 - 0.014\,t_3,
\label{equation:o3}
\end{eqnarray}
and
\begin{eqnarray}
12+ \log (O^{+}/H^+) = \log (I_{[OII] \lambda 3727+ \lambda 3729}/I_{ H_{\beta} })
 +  \nonumber  \\
5.961+ \frac{1.676}{t_2}  - 0.40 \log t_2  -0.034\,t_2 + \log (1+1.35x_2)  ,
\label{equation:o2}
\end{eqnarray}
where
\begin{equation}
x_2= 10^{-4} n_{\rm e} t_2^{-1/2}.
\end{equation}
Here $n_e$ is the electron density in cm$^{-3}$. 

The total oxygen abundances are then derived from the following equation: 
\begin{equation}
\frac{O}{H} = \frac{O^+}{H^+} + \frac{O^{++}}{H^+}                .
\label{equation:otot}
\end{equation}

The electron temperature $t_2$ of the [O\,{\sc ii}] zone is 
usually determined from an equation which relates $t_2$ to $t_3$, 
derived by fitting H\,{\sc ii} region models. 
Several versions of this $t_2$ -- $t_3$ relation have been proposed. 
A widely used relation is the one by \citet{campbelletal86} 
(see also \citet{garnett92}) based on the H\,{\sc ii} region models of 
\citet{stasinska82}:
\begin{equation}
t_2 = 0.7 \, t_3 + 0.3.
\label{equation:ttcam}
\end{equation}
Another relation has been proposed by \citet{pageletal92}, also 
based on H\,{\sc ii} region models of 
\citet{stasinska90})
\begin{equation}
\frac{1}{t_2} = 0.5(\frac{1}{t_3} + 0.8).
\label{equation:ttpag}
\end{equation}
\citet{izotovetal97}, fitting also
the H\,{\sc ii} region models 
of \citet{stasinska90}, have proposed the following expression
\begin{equation}
t_2 =  0.243 + 1.031\,t_3  - 0.184\,t_3^2  .
\label{equation:ttiz}
\end{equation}
Based on H\,{\sc ii} region model calculations by 
\citet{stasinskaschaerer97}, \citet{deharvengetal00} 
have suggested the following relation 
\begin{equation}
t_2 = 0.775 \, t_3 + 0.281.
\label{equation:ttdeh}
\end{equation}
\citet{oeyshields00} have found that the Campbell et al. relation 
is reasonable for $t_{3}$ $>$ 1.0. However at lower temperatures, 
the models are more consistent with an isothermal nebula. They consequently
adopted the formulation, 
\begin{eqnarray}
t_2   = & 0.7 \, t_3 + 0.3, &  \;\; t_3 > 1.0   \nonumber  \\
        & t_3,              &  \;\; t_3 < 1.0 .
\label{equation:ttoey}
\end{eqnarray}

Using the system of equations Eqs.(\ref{equation:t3})$\div$(\ref{equation:otot}) 
and one of the t$_2$ -- t$_3$ relations chosen among  
Eqs.(\ref{equation:ttcam})$\div$(\ref{equation:ttoey})), 
the oxygen abundance (O/H)$_{\rm T_e}$ in an H\,{\sc ii} region 
can then be determined.

\subsection{An alternative version of the T$_{{\rm e}}$ method}

Examination of Eq.(\ref{equation:o3}) shows that the fluxes 
used to calculate R$_3$ and in the
H$_{\beta}$ line originate from different volumes of the nebula. 
The H$_{\beta}$ flux comes from the whole H\,{\sc ii} region while the 
R$_3$ fluxes come from the volume where the oxygen is in the O$^{++}$ stage. 
If Eq.(\ref{equation:o3}) is applied to the O$^{++}$ zone only, then 
it gives the total $\frac{O}{H}$ instead 
of the ionic $\frac{O^{++}}{H^+}$ oxygen abundance as 
in the traditional approach.
In that case, Eq.(\ref{equation:o3}) takes the form 
\begin{eqnarray}
12+ \log (O/H) = \log (I_{[OIII] \lambda 4959+ \lambda 5007}/
(w \times I_{H_{\beta} }))   \nonumber  \\
+ 6.200 + \frac{1.251}{t_3}  - 0.55 \log t_3 - 0.014\,t_3 ,
\label{equation:o3w1}
\end{eqnarray}
where {\it w} is the fraction of the H$_{\beta}$ flux in the O$^{++}$ zone.
With the adopted notations, Eq.(\ref{equation:o3w1}) can be 
rewritten as 
\begin{eqnarray}
12+ \log (O/H) = \log (R_3/w) + 6.200   \nonumber  \\
 + \frac{1.251}{t_3}  - 0.55 \log t_3 - 0.014\,t_3.
\label{equation:o3w}
\end{eqnarray}
Similarly, Eq.(\ref{equation:o2}) can be rewritten as 
\begin{eqnarray}
12+ \log (O/H) = \log (I_{[OII] \lambda 3727+ \lambda 3729}/
[(1-w) \times I_{ H_{\beta} }]   \nonumber  \\
+ 5.961+ \frac{1.676}{t_2}  - 0.40 \log t_2  -0.034\,t_2  \nonumber  \\
+ \log (1+1.35x_2)  .
\label{equation:o2w1}
\end{eqnarray}
or
\begin{eqnarray}
12+ \log (O/H) = \log [R_2/(1-w)] + 5.961+ \frac{1.676}{t_2}  \nonumber  \\
- 0.40 \log t_2  -0.034\,t_2  
+ \log (1+1.35x_2)  .
\label{equation:o2w}
\end{eqnarray}
There is no reason to expect the oxygen abundance in the O$^{++}$ zone to 
differ from that in the O$^{+}$ zone within the same H\,{\sc ii} region. 
In other words, the oxygen abundance derived from Eq.(\ref{equation:o3w}) must 
be equal to the one derived from Eq.(\ref{equation:o2w}). 
Equating the right-hand sides of these two equations result in 
\begin{eqnarray}
\log (R_3/w) + 6.200 
 + \frac{1.251}{t_3}  - 0.55 \log t_3 - 0.014\,t_3 =  \nonumber  \\
 \log [R_2/(1-w)] + 5.961+ \frac{1.676}{t_2} - 0.40 \log t_2   \nonumber  \\
 - 0.034\,t_2  + \log (1+1.35x_2)  .
\label{equation:bas}
\end{eqnarray}
Eq.(\ref{equation:bas}) can be solved for t$_{2}$ if the values of t$_3$, 
n$_{\rm e}$ and {\it w} are known. The majority of extragalactic H\,{\sc ii} 
regions are in the low-density regime \citep{zkh,bresolinetal05}. 
Therefore we adopt 
an electron density n$_{\rm e}$ = 100 cm$^{-3}$ for all the
H\,{\sc ii} regions in our sample. 
Thus, Eq.(\ref{equation:bas}) allows  
to derive the electron temperature $t_2$ within the [O\,{\sc ii}] 
zone in a H\,{\sc ii} region if the electron temperature $t_3$ within the 
[O\,{\sc iii}] zone and the fraction {\it w} of H$_{\beta}$ flux in 
the O$^{++}$ zone are known.
 
\subsection{Estimation of the parameter {\it w}}

\begin{figure}
\resizebox{1.00\hsize}{!}{\includegraphics[angle=000]{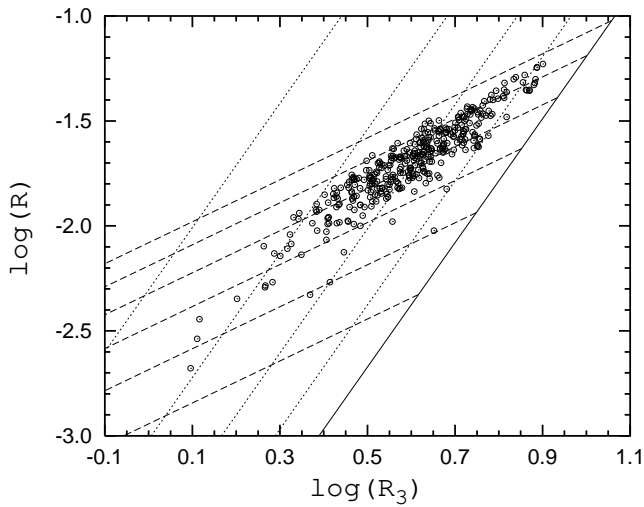}}
\caption{The R vs. R$_{3}$ diagram. 
The circles show individual H\,{\sc ii} region measurements from the  
standard sample. The relation corresponding to Eq.(\ref{equation:ffp}) for 
the value of the excitation parameter P = 1.0 is shown by the solid line. 
The dotted lines show from right to left the relations corresponding to that 
equation for P = 0.8, 0.6, 0.4 and 0.2. The dashed lines show from top 
to bottom the positions of H\,{\sc ii} regions with a given value of the 
electron temperature t$_3$ = 1.2, 1.1, 1.0, 0.9, 0.8 and 0.7, 
in units of 10$^4$~K. }
\label{figure:f-f}
\end{figure}

The fraction {\it w} of H$_{\beta}$ flux in the O$^{++}$ zone can be estimated 
in the following way. We have investigated recently the 
relationship between the observed auroral and nebular 
oxygen line fluxes in spectra of H\,{\sc ii} regions \citep{ff,pilyuginetal06}.
We have found a relation (called the ff relation) 
that is metallicity-dependent at low metallicities, 
but independent of metallicity above 12+logO/H $\sim$ 8.25, i.e. there is 
one-to-one correspondence between the auroral and nebular
oxygen line fluxes in spectra of high-metallicity H\,{\sc ii} regions. 
Using a compilation of recent high-precision 
measurements of oxygen lines fluxes in high-metallicity H\,{\sc ii} 
regions, the following ff relation was derived \citep{pilyuginetal06}
\begin{eqnarray}
\log R & = & - 4.151- 3.118\,\log P + 2.958 \, \log R_{3}
\nonumber  \\
       & - & 0.680 \, (\log P)^2 .
\label{equation:ffp}
\end{eqnarray}
Fig.~\ref{figure:f-f} shows the relation corresponding 
to Eq.(\ref{equation:ffp}) in the R$_3$ -- R diagram for various values of 
the excitation parameter P.
The solid line, called hereafter as the basic line, corresponds 
to P = 1.0, while   
the dotted lines correspond, from right to left, to P = 0.8, 0.6,
0.4, 0.2. The dashed lines show the loci of 
H\,{\sc ii} regions for several values  
of the electron temperature t$_3$. From top to bottom, t$_3$ = 1.2, 
1.1, 1.0, 0.9, 0.8, 0.7 in units of 10$^4$~K. 
The circles show individual measurements of H\,{\sc ii} regions from the 
present sample. 

Let us consider a sequence of H\,{\sc ii} regions with the same value of t$_3$.
The fraction of radiation of the H\,{\sc ii} region in the 
O$^{++}$ zone increases along this sequence from low to high values 
of P, and reaches its maximum value, {\it w} = 1, at 
P = 1, i.e. at the intersection point  
of a dashed line with a given electron temperature  
with the basic solid line . 
Then, the fraction of radiation in the O$^{++}$ zone can be estimated as 
\begin{equation}
 w  = R_{3}^{\rm obs}/R_{3}^{\rm P=1} .
\label{equation:w}
\end{equation}
Such an estimate is not airtight since not only  
{\it w} can change along a 
sequence of H\,{\sc ii} regions with a fixed value of t$_3$, 
but other characteristics of H\,{\sc ii} regions, such as, for example,
their oxygen abundance, can do so as well.
Therefore, the value of t$^{*}_2$ derived from Eq.(\ref{equation:bas}) and 
Eq.(\ref{equation:w}) may carry some systematic error {\it E}$_{\rm t}$. 
That error should be small for high-excitation H\,{\sc ii} 
regions (they are located close to the basic line) but increase with 
decreasing P. In the following, we discuss how to 
estimate that error and correct t$^{*}_2$ for it.

\section{The \lowercase{t$_2$--t$_3$} relation}

\begin{figure}
\resizebox{1.00\hsize}{!}{\includegraphics[angle=000]{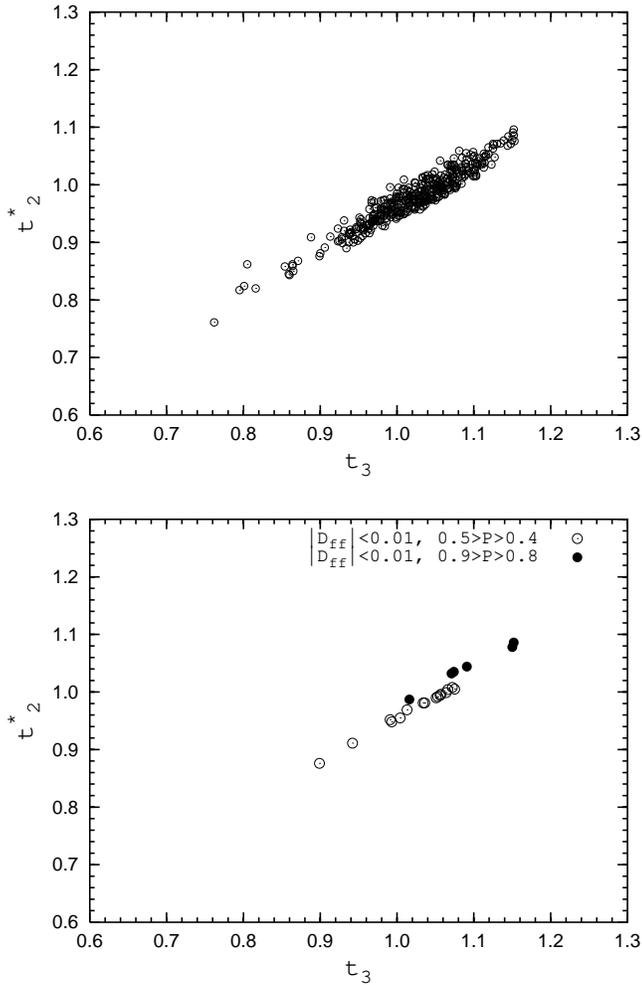}}
\caption{
{\it Top panel.} 
The electron temperatures t$^*_2$ versus t$_3$ for our standard 
sample of H\,{\sc ii} regions. 
{\it Bottom panel.} 
The same as in the top panel, except that only high-quality data with 
$|$D$_{\rm ff}|$ $<$ 0.01 are included.
Filled circles show 
H\,{\sc ii} regions with 0.9 $>$ P $>$ 0.8, and open circles 
those with 0.5 $>$ P $>$ 0.4.   }
\label{figure:t2-t3}
\end{figure}

\begin{figure}
\resizebox{1.00\hsize}{!}{\includegraphics[angle=000]{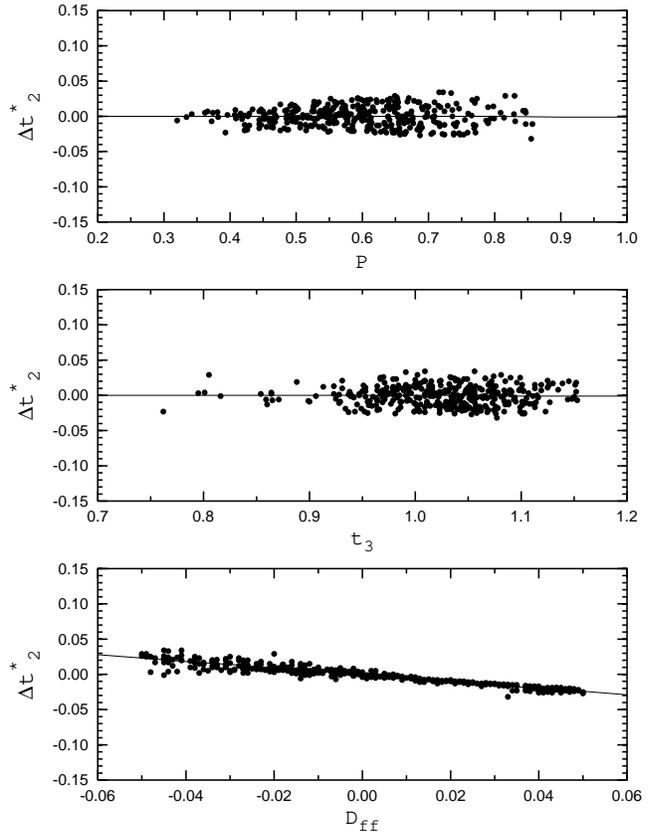}}
\caption{The difference $\Delta$t$^*_2$, defined in section 4,  
as a function of excitation parameter 
({\it top panel}), electron temperature t$_3$ ({\it middle panel}), 
and discrepancy index D$_{\rm ff}$ ({\it bottom panel})  
for our standard sample of H\,{\sc ii} regions. }
\label{figure:dt2}
\end{figure}

The top panel in Fig.~\ref{figure:t2-t3} shows the value of t$^{*}_2$ as 
derived 
from Eq.(\ref{equation:bas}) and Eq.(\ref{equation:w}) for the objects 
in our standard
subsample of high-metallicity H\,{\sc ii} regions. 
As noted above, t$^{*}_2$ has a systematic error {\it E}$_{\rm t}$ that 
depends on the value of the excitation parameter P. To illustrate the point, 
we show in the bottom panel of Fig.~\ref{figure:t2-t3} 
high quality data (those with $|$D$_{\rm ff}|<$ 0.01) for 
H\,{\sc ii} regions with 0.9 $>$ P $>$ 0.8 (filled circles) and for those 
with 0.5 $>$ P $>$ 0.4 (open circles).
Inspection of the bottom panel in Fig.~\ref{figure:t2-t3} shows 
that indeed the t$^{*}_2$ -- t$_3$ relations are slightly different for 
H\,{\sc ii} regions with different values of P: at a given value of t$_3$,
t$^{*}_2$ = t$_2$ + E$_t$ is slightly smaller, on average, 
for low than for high excitation H\,{\sc ii} regions. 
The true t$_2$ -- t$_3$ relation can be found by extrapolation of the 
t$^{*}_2$ -- t$_3$ relation to P = 1.

The bottom panel in Fig.~\ref{figure:t2-t3} shows that the 
systematic error {\it E}$_{\rm t}$ is small. Therefore, we will approximate 
it by a linear expression, i.e.
the true t$_2$ and the derived t$^{*}_2$ values are related by
\begin{equation}
 t_2  = t^*_2 - E_t = t^*_2 - c\;(1 - P) .
\label{equation:e1}
\end{equation}
Examination of Fig.~\ref{figure:t2-t3} shows also that the $t_2$ -- $t_3$  
relation can be parametrized as
\begin{equation}
t_2 = a\, t_3 + b .
\label{equation:tt1}
\end{equation}
The coefficients {\it a} and {\it b} in Eq.(\ref{equation:tt1}) and 
the coefficient {\it c} in Eq.(\ref{equation:e1}) can be found simultaneosly 
by fitting the data 
in the top panel of Fig.~\ref{figure:t2-t3}) by an expression of 
the type
\begin{equation}
 t^*_2  = a\, t_3 + b  + c\;(1 - P) .
\label{equation:e}
\end{equation}
Carrying out a standard least-squares fit to the data, we obtain 
the following values of the 
coefficients;
{\it a} = 0.716 $\pm$ 0.012,
{\it b} = 0.264 $\pm$ 0.013,
{\it c} = --0.042 $\pm$ 0.007. 
Then, the true t$_2$ -- t$_3$ relation is 
\begin{equation}
t_2 = 0.716 (\pm 0.012) \, t_3 + 0.264 (\pm 0.013).
\label{equation:tt}
\end{equation}

The value of {\it w} given by Eq.(\ref{equation:w}) is the fraction of 
radiation 
in the O$^{++}$ zone if the condition t$_2$=t$_3$ holds. 
This is not however the case. 
Since the H$_{\beta}$ emission coefficient, E$^{0}_{4,2}$, 
is dependent on the
electron temperature \citep{aller84}
\begin{equation}
E^0_{4,2}  = 1.387\, t^{-0.983} \;10^{-0.0424/t} ,
\label{equation:e42}
\end{equation}
then the contribution {\it w}$_{\rm c}$ 
of the [OIII] zone to the total flux of the nebula in the 
H$_{\beta}$ line is given by the
following expression
\begin{equation}
{\it w}_{\rm c} = \frac{E^0_{4,2}(t_3)\;{\it w}}
{E^0_{4,2}(t_3)\;{\it w} + E^0_{4,2}(t_2)\;(1 - {\it w})}  .
\label{equation:wc}
\end{equation}
We have recomputed the values of t$^{*}_2$ using 
{\it w}$_{\rm c}$ instead 
of {\it w}. We found the differences between the t$^{*}_2$ values derived with 
{\it w} and {\it w}$_{\rm c}$ to be less than 0.01. Therefore, 
we will neglect this small correction, and will use 
the value of {\it w} given by 
Eq.(\ref{equation:w}).

We now check whether the linear forms of the analytical expressions 
adopted for the t$_2$ -- t$_3$ and t$^*_2$ -- t$_2$ relations
are justified. 
The top panel of Fig.~\ref{figure:dt2} shows the difference 
$\Delta$t$^*_2$ between the value of t$^*_2$ derived 
from Eq.(\ref{equation:bas}) and Eq.(\ref{equation:w}) and the one derived 
from Eq.(\ref{equation:e}) as a function of the excitation parameter P.
The points are individual measurements of H\,{\sc ii} regions from the 
standard subsample, and the line 
is the linear best fit to those data obtained through the least-squares method.
The middle panel in Fig.~\ref{figure:dt2} shows the same difference 
$\Delta$t$^*_2$ as a function of the electron temperature t$_3$. 
Examination of the top and middle panels of Fig.~\ref{figure:dt2} shows 
that $\Delta$t$^*_2$ does not coorelate either with
P or t$_3$, justifying our adoption of linear forms
for the t$_2$ -- t$_3$ and t$^*_2$ -- t$_2$ relations. 

The bottom panel in Fig.~\ref{figure:dt2} shows 
$\Delta$t$^*_2$ as a function of the discrepancy index  D$_{\rm ff}$. 
There is an anticorrelation of the two quantities. 
Since the discrepancy index appears to be an indicator of the error 
in the auroral line R measurements,  
this suggests that the t$_2$ -- t$_3$ correlation is rather tight, and that  
any scatter in this correlation is caused mainly by uncertainties in the
measurements.

\begin{figure}
\resizebox{1.00\hsize}{!}{\includegraphics[angle=000]{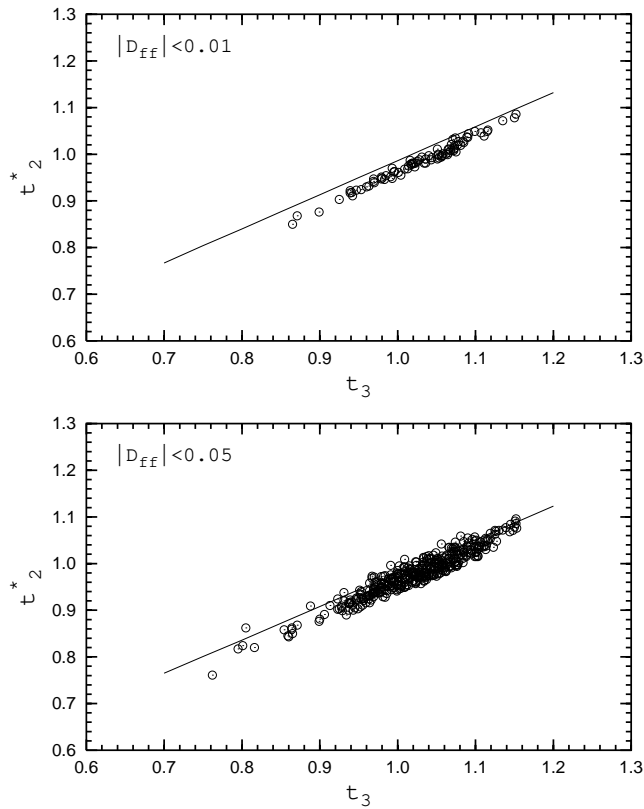}}
\caption{
Open circles  show the t$^*_2$ -- t$_3$ diagram for the 
subsample of H\,{\sc ii} 
regions with  $|$D$_{\rm ff}|$ $<$ 0.01 ({\it top panel}), and for the 
standard sample ({\it bottom panel}).
}
\label{figure:t2-t3-df}
\end{figure}

\begin{figure}
\resizebox{1.00\hsize}{!}{\includegraphics[angle=000]{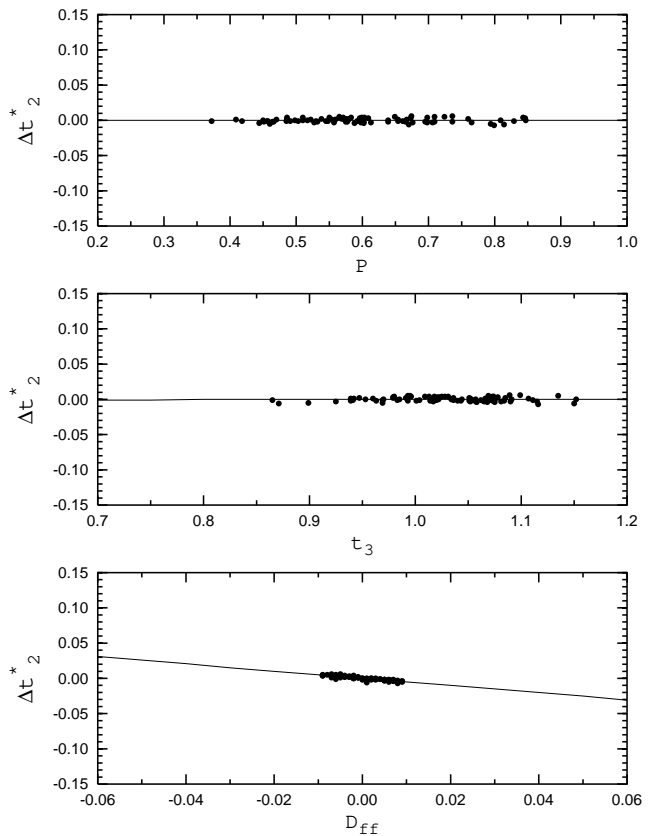}}
\caption{The difference $\Delta$t$^*_2$, defined in section 4, 
as a function of excitation parameter P
({\it top panel}), electron temperature t$_3$ ({\it middle panel}), 
and discrepancy index D$_{\rm ff}$ ({\it bottom panel})  
for the subsample of H\,{\sc ii} regions with $|$D$_{\rm ff}|$ $<$ 0.01. 
}
\label{figure:dt2dff}
\end{figure}

The low uncertainty in the values of the coefficients in Eq.(\ref{equation:tt})
may be a consequence of 
the large number of points (372) used. The following consideration 
can tell us something about the true accuracy of the derived t$_2$ -- t$_3$ 
relation. 
We have extracted from our standard sample 
a subsample of 86 high-quality 
measurements of H\,{\sc ii} regions with $|$D$_{\rm ff}|$ $<$ 0.01. 
The derived t$^*_2$ -- t$_3$ diagram for 
this subsample is shown in the top panel of 
Fig.~\ref{figure:t2-t3-df}. A fit to those data gives 
\begin{eqnarray}
 t^*_2  = 0.729 \, (\pm 0.006) \, t_3 + 0.257 \,(\pm 0.007)     \nonumber  \\
- 0.058 \, (\pm 0.003) \, (1 - P) 
\label{equation:tte01}
\end{eqnarray}
The corresponding t$_2$ -- t$_3$ relation is then
\begin{equation}
 t_2  = 0.729 \, (\pm 0.006) \, t_3 + 0.257 \,(\pm 0.007) .  
\label{equation:tt01}
\end{equation}
This relation is shown by the solid line in the top panel of 
Fig.~\ref{figure:t2-t3-df}. 
For comparison, the bottom panel of Fig.~\ref{figure:t2-t3-df} 
shows the t$^*_2$ -- t$_3$ 
diagram and the t$_2$ -- t$_3$ relation for the whole standard sample.
 
Fig.~\ref{figure:dt2dff} shows the differences $\Delta$t$^*_2$ 
between the values of t$^*_2$ derived from Eq.(\ref{equation:bas}) 
and Eq.(\ref{equation:w}) and the one derived from
Eq.(\ref{equation:tte01}) as a function of P
(top panel), t$_3$ (middle panel) and D$_{\rm ff}$ (bottom panel). 
The points are individual H\,{\sc ii} regions with 0.01 $>$ D$_{\rm ff}$ 
$>$ --0.01, the lines are linear least-square fits to those data. 
Comparison of Fig.~\ref{figure:dt2} and Fig.~\ref{figure:dt2dff} shows 
clearly that the scatter in t$_2$ at a fixed value of t$_3$ decreases in the 
subsample of H\,{\sc ii} regions with more precise measurements. This 
confirms our conclusion that the t$_2$ -- t$_3$ correlation is rather tight, 
and that the scatter in it is caused mainly by measurement uncertainties.

Examination of Eq.(\ref{equation:tt}) and  Eq.(\ref{equation:tt01}) shows 
that the t$_2$ -- t$_3$ relations derived from the 
two different samples of H\,{\sc ii} regions
are very similar and agree within the formal uncertainties. Thus the 
derived t$_2$ -- t$_3$ relation is rather robust. In the following,
we will adopt as the t$_2$ -- t$_3$ relation
\begin{equation}
 t_2  = 0.72 \, t_3 + 0.26  
\label{equation:ttfin}
\end{equation}

We note that 
it has been generally accepted that there is a one-to-one 
correspondence between t$_2$ and t$_3$, i.e. that the 
t$_2$ -- t$_3$ relation does not depend on an additional parameter. 
If there is a dependence of the t$_2$ -- t$_3$ relation  
on the excitation parameter (from general considerations, this 
cannot be excluded), that dependence cannot be revealed by our 
approach and can influence the derived relation. 

In summary, there is a tight correlation between the electron temperature $t_3$
within the [O\,{\sc iii}] zone and the electron temperature $t_2$ within the 
[O\,{\sc ii}] zone in high-metallicity H\,{\sc ii} regions. 
This correlation can be well approximated by a linear expression and its 
form is rather robust.

\section{Discussion}

\begin{figure}
\resizebox{1.00\hsize}{!}{\includegraphics[angle=000]{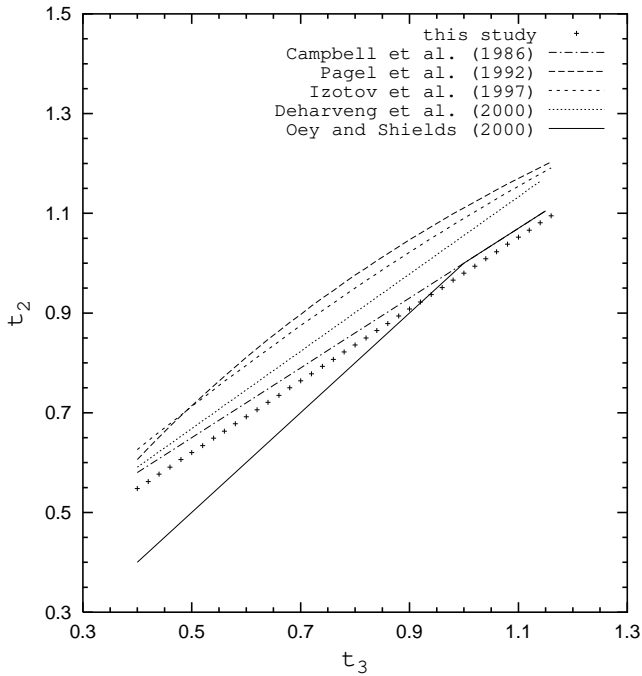}}
\caption{Comparison of the t$_2$ -- t$_3$ relation derived here 
with those derived by 
other investigators.}
\label{figure:compar}
\end{figure}

\begin{figure}
\resizebox{1.00\hsize}{!}{\includegraphics[angle=000]{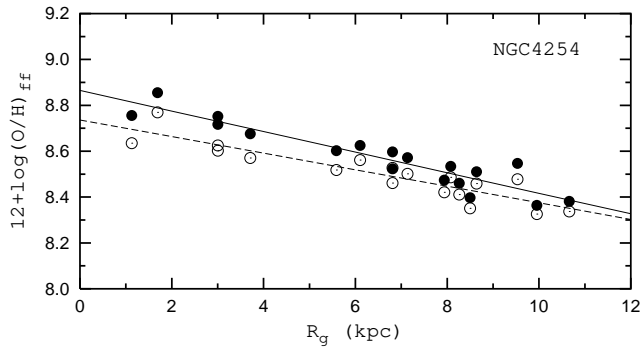}}
\caption{The radial distribution of oxygen abundances across the disk of the 
spiral galaxy NGC~4254. The filled circles show (O/H)$_{\rm ff}$ abundances 
based on the t$_2$ -- t$_3$ relation derived in the present study, and 
the solid 
line is the linear least-squares best fit to those data. 
The open circles show 
(O/H)$_{\rm ff}$ abundances based on the t$_2$ -- t$_3$ 
relation of \citet{campbelletal86}, and 
the dashed line is the linear least-squares best fit 
to those data. }
\label{figure:ngc4254}
\end{figure}

We now compare the t$_2$ -- t$_3$ relation obtained here with those 
obtained 
by other authors. Fig.~\ref{figure:compar} shows our t$_2$ -- t$_3$ relation 
together with those from 
\citet{campbelletal86,pageletal92,izotovetal97,deharvengetal00,oeyshields00}. 
Since our t$_2$ -- t$_3$ relation is derived for 
cool high-metallicity H\,{\sc ii} regions then the high-temperature 
low-metallicity
part of relation is not considered here. 
Fig.~\ref{figure:compar} shows that our t$_2$ -- t$_3$ relation 
is most similar to the one by \citet{campbelletal86}. 

We consider next how the obtained t$_2$ -- t$_3$ relation may affect the 
derived oxygen abundances in H\,{\sc ii} regions. 
Since the t$_2$ -- t$_3$ relation of \citet{campbelletal86}
has found wide acceptance and use in abundance determinations in H\,{\sc ii} 
regions, we will compare oxygen abundances derived with 
the \citet{campbelletal86} relation and ours. For this purpose, we use the data
for H\,{\sc ii} regions in the spiral galaxy NGC~4254 obtained by 
\citet{mrs,shieldsetal91,henryetal94}.
 Since the auroral 
[O\,{\sc iii}]$\lambda 4363$ line was not detected in those H\,{\sc ii} 
regions, O/H)$_{\rm ff}$ abundances based on the line fluxes 
R$^{\rm cal}$ have been 
determined from the ff relation, Eq.(\ref{equation:ffp}), 
for every H\,{\sc ii} region, following \citet{pilyuginetal06}. 
 The (O/H)$_{{\rm ff}}$ abundances in 
the H\,{\sc ii} regions of NGC~4254 determined using the t$_2$ -- t$_3$ 
relation of \citet{campbelletal86} are shown as a function of galactocentric 
distance in Fig.~\ref{figure:ngc4254} by open circles. 
The dashed line is the linear least-squares best fit to those data:  
\begin{equation}
12+\log(O/H) = 8.74\,(\pm 0.03) - 0.036\, (\pm 0.004) \times R_{\rm g} .
\label{equation:ngccam}
\end{equation}
We have adopted a distance of 16.14 Mpc for NGC~4254 \citep{pilvilcon04}. 
The filled circles are (O/H)$_{{\rm ff}}$ 
abundances determined with our t$_2$ -- t$_3$ relation. 
The solid line is the linear least-squares best fit to those data:  
\begin{equation}
12+\log(O/H) = 8.86\,(\pm 0.03) - 0.045\, (\pm 0.004) \times R_{\rm g} .
\label{equation:ngcpil}
\end{equation}
Fig.~\ref{figure:ngc4254} shows that the abundances derived with our 
t$_2$ -- t$_3$ relation 
are slightly higher (up to $\sim$ 0.1 dex) than those derived 
with the relation by \citet{campbelletal86}. Comparison of 
Eq.(\ref{equation:ngccam}) and Eq.(\ref{equation:ngcpil}) leads to the same 
conclusion.

\section{Conclusions}

We suggest a new way to establish the 
relation between the electron temperature 
$t_3$ within the [O\,{\sc iii}] zone and the electron temperature $t_2$ within 
the [O\,{\sc ii}] zone in high-metallicity (12+log(O/H) $>$ 8.25) H\,{\sc ii} 
regions. The basic idea is the following. If we apply the equation used  
to calculate the ionic abundance  
O$^{++}$/H$^+$ not to the entire H\,{\sc ii} region but 
only to the O$^{++}$ zone, then this would yield, not the ionic 
O$^{++}$/H$^+$ abundance but the total O/H oxygen abundance instead. 
We require that the equation for O$^{++}$/H$^+$ applied to the O$^{++}$ zone 
and the one for O$^{+}$/H$^+$ applied to the O$^{+}$ zone result in exactly 
the same value of the oxygen abundance. This condition allows us to derive a 
relation between t$_2$ and t$_3$. We have applied this method to a sample 
of 372 H\,{\sc ii} regions selected to have high-quality measurements by using 
the ff relation \citep{pilyuginetal06}. We find that the correlation between 
t$_2$ and t$_3$ is tight and can be approximated by a linear expression. 
The so derived $t_2$ -- $t_3$ relation is 
independent of photoionization models of H\,{\sc ii} regions.

The derived relation can be used to determine t$_2$ and accurate abundances in 
high-metallicity H\,{\sc ii} regions with a measured t$_3$ temperature. 
It can be also used in conjunction with the ff relation 
of \citet{pilyuginetal06} for the determination of the t$_3$ and t$_2$ 
temperatures and oxygen abundances in high-metallicity H\,{\sc ii} regions 
where the [OIII]$\lambda$4363 auroral line is not detected. 

Our t$_2$ -- t$_3$ relation is close to the widely used relation of 
\citet{campbelletal86}. However, the abundances derived with our 
t$_2$ -- t$_3$ relation are slightly higher (up to $\sim$ 0.1 dex) than those 
derived with the \citet{campbelletal86} relation.

\subsection*{Acknowledgments}

   We thank Yuri Izotov and Natalia Guseva for providing us with their 
measurements of the line intensities in spectra of H\,{\sc ii} regions extracted 
from the Data Release 3 of the Sloan Digital Sky Survey (SDSS), the majority of 
which are not published. 
We are grateful to Bernard Pagel for a numerous constructive comments 
over all the period of the performance of this study. 
We thank the anonymous referee for helpful comments. 
   This research was made possible in part by Award No. UP1-2551-KV-03 of the US 
Civilian Research \& Development Foundation for the Independent States of the 
Former Soviet Union (CRDF). 
   T.X.T. has been partially supported by NSF grant AST-02-05785. T.X.T. thanks 
the hospitality of the Institut d'Astrophysique in Paris and of the Service 
d'Astrophysique at Saclay during his sabbatical leave. He is grateful for a 
Sesquicentennial Fellowship from the University of Virginia. 
   All the authors acknowledge the work of the SDSS team. 
Funding for the SDSS has been provided by the
Alfred P. Sloan Foundation, the Participating Institutions, the National
Aeronautics and Space Administration, the National Science Foundation, the
U.S. Department of Energy, the Japanese Monbukagakusho, and the Max Planck
Society. The SDSS Web site is http://www.sdss.org/.
     The SDSS is managed by the Astrophysical Research Consortium (ARC) for
the Participating Institutions. The Participating Institutions are The
University of Chicago, Fermilab, the Institute for Advanced Study, the Japan
Participation Group, The Johns Hopkins University, the Korean Scientist Group,
Los Alamos National Laboratory, the Max-Planck-Institute for Astronomy (MPIA),
the Max-Planck-Institute for Astrophysics (MPA), New Mexico State University,
University of Pittsburgh, University of Portsmouth, Princeton University, the
United States Naval Observatory, and the University of Washington.

\end{document}